\documentclass[11pt]{article}

\usepackage[letterpaper,margin=1in]{geometry}
\usepackage[T1]{fontenc}
\usepackage[utf8]{inputenc}
\usepackage{lmodern}
\usepackage{microtype}
\usepackage{amsmath,amssymb}
\usepackage{graphicx}
\usepackage{booktabs}
\usepackage{array}
\usepackage{algorithm}
\usepackage{algpseudocode}
\usepackage{caption}
\usepackage{float}
\usepackage{placeins}
\usepackage{cuted}
\usepackage[numbers,sort&compress]{natbib}
\usepackage[hyphens]{url}
\usepackage[colorlinks=true,linkcolor=blue,citecolor=blue,urlcolor=blue]{hyperref}

\newcommand{\MusicLayout}{\textsc{MusicLayout}}

\title{MusicLayout: Explicit Structural Planning\\ for Controllable Text-to-Music Generation}

\author{%
Shuyu Li$^{1}$ \and
Kejun Zhang$^{1,3}$\thanks{Corresponding author.} \and
Jiahe Lei$^{4}$ \and
Shulei Ji$^{2,3}$ \and
Zihao Wang$^{2,5}$ \and
Jiaxing Yu$^{1}$ \and
Wanying Wu$^{6}$ \and
Lei Wang$^{7}$\\[0.5em]
\small $^{1}$College of Artificial Intelligence, Zhejiang University\\
\small $^{2}$College of Computer Science and Technology, Zhejiang University\\
\small $^{3}$Innovation Center of Yangtze River Delta, Zhejiang University\\
\small $^{4}$The Chinese University of Hong Kong\\
\small $^{5}$Shandong University\\
\small $^{6}$Chu Kochen Honors College, Zhejiang University\\
\small $^{7}$Ant Group\\[0.5em]
\small \{lsyxary, zhangkejun, shuleiji, yujx, 3240100265\}@zju.edu.cn\\
\small 1155261729@ee.cuhk.edu.hk, carlwang1212@gmail.com, thirtyking@163.com
}

\date{}

\begin{document}

\maketitle

\begin{abstract}
Text-to-music generation has advanced rapidly, but current systems still rely primarily on global text prompts, leaving the structural organization of generated music implicit and difficult to inspect, control, or revise before audio generation. To address this issue, we introduce MusicLayout, an explicit intermediate representation for controlling musical structure in text-to-music generation. MusicLayout describes a musical piece as a time-aligned layout of sections, textures, repetitions, variations, and instrument-level arrangements, serving as an interpretable planning layer between textual intent and the generated music. We integrate MusicLayout into a text-to-music framework built on a unified autoregressive formulation, where the model first generates a MusicLayout representation and subsequently predicts audio tokens conditioned on this representation within a single sequence. The resulting MusicLayout can be inspected and modified prior to audio generation, providing a mechanism for layout-level structural control. We evaluate MusicLayout through layout-conditioned generation, layout manipulation experiments, and matched-data ablations, providing evidence that explicit layout planning can improve long-range structural organization and support layout-level control. We have released the implementation as open source on GitHub at \url{https://github.com/XaryLee/MusicLayout}.
\end{abstract}

\section{Introduction}

Recent text-to-music models have advanced substantially in fidelity,
semantic alignment, musicality, and efficiency~\cite{li2026survey}. They
typically generate discrete audio tokens autoregressively or synthesize
continuous and compressed representations with diffusion models
\cite{agostinelli2023musiclm,copet2024musicgen,liu2024audioldm2}. Recent
systems further support long-form generation and combine language-model (LM)
reasoning with diffusion-based acoustic synthesis
\cite{evans2026stable,gong2026ace}.

Despite these advances, most systems rely on global text prompts and leave
musical planning implicit. A prompt can describe genre, mood,
instrumentation, or tempo, but cannot precisely specify how sections are
organized, materials recur or vary, textures evolve, or instruments enter
and leave. Existing controllable methods add melody, chords, drums,
dynamics, symbolic lead sheets, or semantic representations
\cite{wu2024music,DITTO,tal2024jasco,bai2024seedmusic,gong2026ace}. These
controls are often limited to individual attributes, require additional
musical inputs, or remain implicit. They do not expose a unified,
time-aligned representation for specifying the organization and arrangement
of an entire piece.

Explicit planning representations improve controllability and
interpretability in image generation by exposing high-level organization
before content. PlanGen~\cite{he2025plangen}, for example, generates spatial
layout tokens before image tokens in one autoregressive sequence. For music,
this principle could improve long-range structural organization and
controllability. Transferring it requires representing organization along
time rather than space.

\begin{figure*}[t]
\centering
\includegraphics[width=0.98\textwidth]{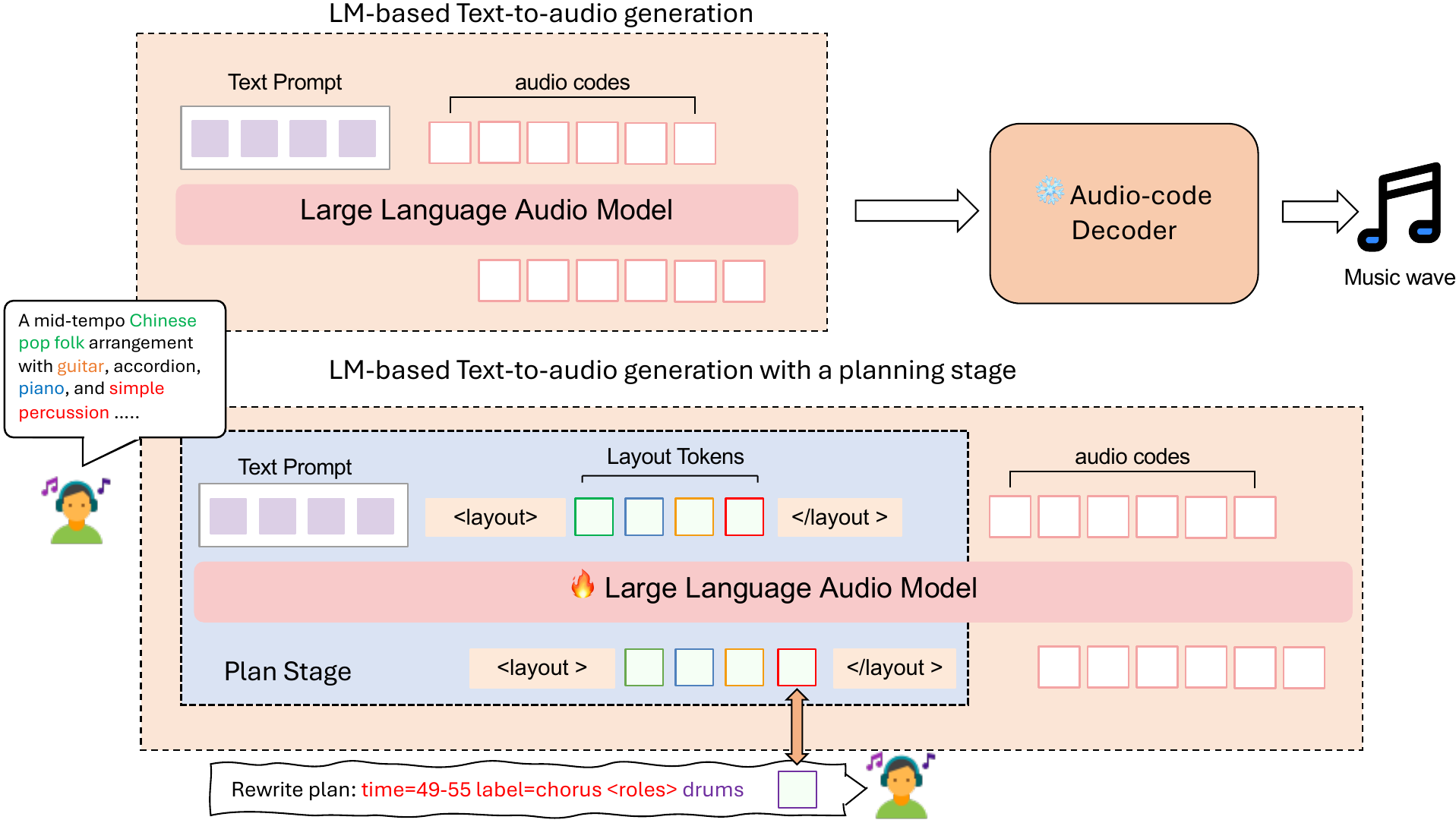}
\caption{Conventional autoregressive LM-based text-to-audio generation (top)
and our explicit-planning framework (bottom). Our LM generates MusicLayout
tokens and then discrete audio tokens, labeled audio codes in the figure,
within one sequence. A frozen decoder renders the audio tokens into music.
The abbreviated rewrite selects the musical section at 49--55 seconds,
assigns it the \texttt{chorus} label, and sets its active instrument list to
drums. The \texttt{<roles>} tag delimits that list.}
\label{fig:framework}
\end{figure*}

Focusing on instrumental music, we introduce
\textbf{MusicLayout}, an explicit intermediate representation for planning
musical structure in audio language models. It describes
section transitions, repetitions and variations, texture changes, and
instrument participation along a shared timeline. As illustrated in
Figure~\ref{fig:framework}, given a text prompt, our model generates a
MusicLayout representation and subsequently predicts audio tokens conditioned
on it within a single sequence. This makes structural
planning inspectable before synthesis and supplies audio generation with an
explicit description of how the music should unfold. Experiments on
layout conditioning and manipulation provide evidence that MusicLayout can
improve structural organization and support layout-level control.

Our main contributions are summarized as follows:

\begin{itemize}

    \item We propose \textbf{MusicLayout}, an explicit planning
    representation for unified audio language models. It represents music
    as a time-aligned layout of sections, repetitions, variations,
    textures, and instrument arrangements, providing structured musical
    plans beyond global textual prompts.

    \item We develop a unified autoregressive framework for
    \textbf{layout planning and audio generation}, where a single model
    first produces an explicit MusicLayout plan and then continues to
    generate audio tokens conditioned on the preceding layout within a single
    sequence. By making musical planning explicit, this framework
    supports long-range structural organization and layout-level control.

    \item Through evaluations of \textbf{layout-conditioned generation,
    layout manipulation, and matched-data ablations}, we provide evidence
    that MusicLayout can serve as an interpretable planning interface and
    improve long-range structural organization in text-to-music generation.

\end{itemize}

\section{Related Work}

\subsection{Text-to-Music Generation}

Autoregressive music models predict discrete audio tokens, as exemplified by
Jukebox, MusicLM, and MusicGen
\cite{dhariwal2020jukebox,agostinelli2023musiclm,copet2024musicgen}.
Diffusion-based systems instead synthesize waveforms, spectrograms, or
compressed latents through iterative denoising
\cite{huang2023noise2music,mousai-2024-efficient,melechovsky2024mustango}.
AudioLDM~2 and MeLoDy combine LM-based semantic modeling with diffusion
synthesis~\cite{liu2024audioldm2,lam2024MeLoDy}. Stable Audio~3 supports
efficient variable-length generation, while ACE-Step~1.5 uses a hybrid
architecture in which an LM performs high-level planning and a Diffusion
Transformer (DiT) realizes the audio~\cite{evans2026stable,gong2026ace,peebles2023dit}. Although
some systems incorporate high-level planning, temporal organization and
arrangement remain implicit or coarsely represented.

\subsection{Controllable Text-to-Music Generation}

Prior systems supplement global prompts with melody references, rhythm,
dynamics, chords, drums, or other symbolic and audio conditions
\cite{agostinelli2023musiclm,copet2024musicgen,wu2024music,DITTO,tal2024jasco}.
Seed-Music uses a symbolic lead-sheet pipeline \cite{bai2024seedmusic}. ACE-Step~1.5 uses metadata and
song-blueprint planning, but its blueprint remains implicit in the learned
audio-token representations and is not exposed as an interpretable structural plan
\cite{gong2026ace}. Other methods use individual controls,
additional inputs, task-specific modules, or coarse metadata. MusicLayout
instead exposes sections, textures, repetitions, variations, and instrument
arrangements in an inspectable and adjustable time-aligned plan.

\section{Methodology}

We build our model upon ACE-Step~1.5, extending its LM to generate a
MusicLayout before producing layout-conditioned audio tokens. The original
audio tokenizer and DiT remain frozen, with the latter rendering the tokens
into a waveform.

\begin{figure*}[t]
\centering
\includegraphics[width=\textwidth]{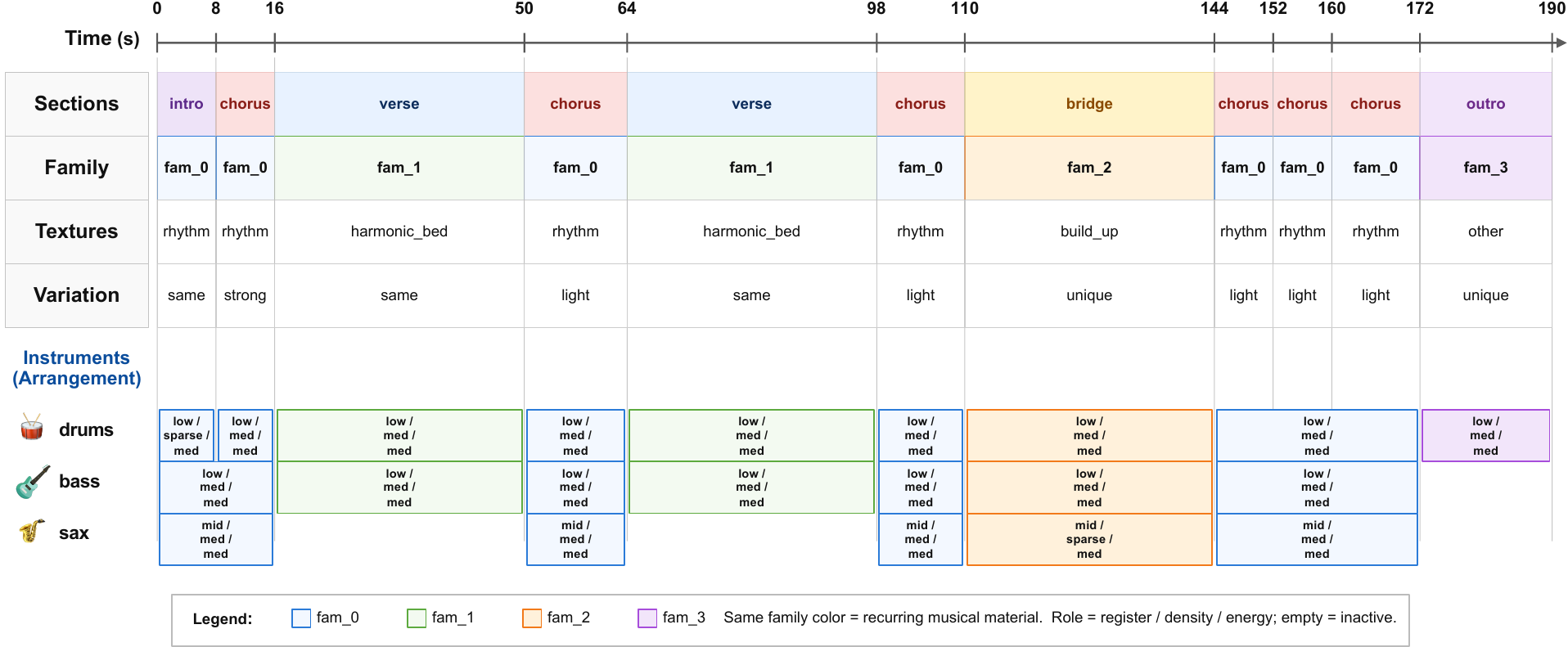}
\caption{A MusicLayout example aligned to a shared timeline. Rows show sections,
material families, textures, variations, and instrument arrangements. Family
colors identify recurring material, and instrument entries give register,
density, and energy. Blanks denote inactivity. For readability, \texttt{rhythm}, \texttt{light}, and
\texttt{strong} abbreviate \texttt{rhythm\_driven}, \texttt{light\_var}, and
\texttt{strong\_var}, respectively.}
\label{fig:musiclayout_example}
\end{figure*}

\subsection{MusicLayout Representation}

MusicLayout externalizes temporal musical organization as a structured,
human-readable sequence that bridges high-level structural intent and
audio-token generation. It follows three design principles: temporal
alignment, hierarchical organization, and controllability.

A MusicLayout combines piece-level structural relationships among musical
materials with time-aligned segments, each corresponding to one musical
section.
At the piece level, MusicLayout organizes recurring musical materials
and their relationships, including repetitions and variations across
sections. At the segment level, each segment is associated with a
time-aligned span and describes its section identity, texture evolution,
and instrument participation.
Its functional section label is selected from \texttt{intro}, \texttt{verse},
\texttt{prechorus}, \texttt{chorus}, \texttt{bridge}, \texttt{breakdown},
\texttt{outro}, \texttt{transition}, \texttt{hook}, \texttt{solo}, and
\texttt{build}.
Figure~\ref{fig:musiclayout_example} illustrates these attributes along a
shared timeline.
MusicLayout is serialized as discrete tokens, allowing it to be directly
modeled by an autoregressive LM.

A \emph{family} is a piece-local group of segments that share similar musical
material. Labels such as \texttt{fam\_0} and \texttt{fam\_1} identify families
only within the current piece. The same label may denote different musical
material in another piece. Each segment is marked
as \texttt{unique}, \texttt{same}, \texttt{light\_var}, or
\texttt{strong\_var}. These values indicate, respectively, that the segment
has no repeated family, closely repeats its family pattern, or departs from
that pattern to a smaller or larger degree. Thus, the family colors and the
variation row in Figure~\ref{fig:musiclayout_example} jointly describe which
material returns and how much it changes.

\begin{table}[t]
\centering
{
\small
\setlength{\tabcolsep}{3pt}
\begin{tabular}{@{}p{0.31\linewidth}p{0.63\linewidth}@{}}
\toprule
Category & Meaning \\
\midrule
\texttt{layered} & Coexisting lead and supporting layers \\
\texttt{rhythm\_driven} & Rhythm-section-centered arrangement \\
\texttt{melodic\_front} & Foregrounded melodic material \\
\texttt{harmonic\_bed} & Sustained harmonic support \\
\texttt{percussive} & Percussion-centered material \\
\texttt{build\_up} & Accumulating or intensifying layers \\
\texttt{sparse\_pulse} & Sparse intermittent activity \\
\texttt{lead\_front} & Foregrounded lead material \\
\texttt{contrast} & Deliberately contrasting texture \\
\texttt{other} & Cases outside the listed categories \\
\bottomrule
\end{tabular}
}
\caption{Texture categories in MusicLayout.}
\label{tab:texture_vocabulary}
\end{table}

The \emph{texture} field gives a coarse description of how a musical segment is
organized. Table~\ref{tab:texture_vocabulary} defines its vocabulary. Each
segment contains a \texttt{<roles>} block listing its active instruments.
Figures~\ref{fig:framework} and~\ref{fig:musiclayout_example} use the
form \texttt{instrument:\allowbreak register:\allowbreak density:\allowbreak energy},
for example \texttt{drums:\allowbreak low:\allowbreak sparse:\allowbreak med}.
The instrument field uses a predefined vocabulary of 25 categories. Each
instrument is followed by register
(\texttt{low}/\texttt{mid}/\texttt{high}), density
(\texttt{sparse}/\texttt{med}/\texttt{dense}), and energy
(\texttt{low}/\texttt{med}/\texttt{high}), encoding pitch range, activity, and
intensity, respectively.
Figure~\ref{fig:framework} abbreviates the edited field as \texttt{drums}.

We derived MusicLayout annotations from symbolic representations by extracting
temporal sections, repetitions, variations, textures, and instrument
arrangements. We rendered the symbolic music into audio and aligned each layout
with its audio and text prompt to form prompt--layout--audio training triples.

\subsection{Unified Layout Planning and Audio Generation}

\textbf{Modeling.}
Given conditioning context $c$, the LM serializes MusicLayout sequence $l$
before audio-token sequence $a$ within a single autoregressive sequence:
\begin{equation}
 y=[c,\texttt{<layout>},l,\texttt{</layout>},a].
\end{equation}
where \texttt{<layout>} and \texttt{</layout>} delimit the planning region.
This ordering makes layout generation precede and condition audio-token
generation. The corresponding joint distribution factorizes as
\begin{equation}
 p(l,a\mid c)=p(l\mid c)p(a\mid c,l).
\end{equation}
Here, $p(l\mid c)$ describes layout planning from the conditioning context,
while $p(a\mid c,l)$ describes audio-token generation given the context and
layout.

\textbf{Training.}
We train the LM on two complementary next-token tasks. Layout planning trains
the model to generate $l$ from $c$. Layout-conditioned audio generation trains
it to predict $a$ from $c$ and a ground-truth (GT) MusicLayout:
\begin{equation}
 \begin{array}{rcl}
 \mathcal{L}_{\mathrm{plan}} &=&-\sum_t \log p(l_t \mid c,l_{<t}), \\
 \mathcal{L}_{\mathrm{audio}} &=&-\sum_i \log p(a_i \mid c,l,a_{<i}).
 \end{array}
\end{equation}
We initially optimize layout planning to learn the MusicLayout schema and the
schema-specific special tokens used to mark its fields and boundaries. We then
alternate the tasks, applying the loss only to the
corresponding layout or audio target span. Only the LM is updated. The
pretrained audio tokenizer and DiT synthesis components remain frozen.

\textbf{Inference and control.}
In automatic inference, the LM generates a MusicLayout from the conditioning
context and then continues with audio tokens. Alternatively, the human-readable
layout can be inspected and adjusted before audio generation to change temporal
organization or instrument arrangement. The revised layout and original
context form the autoregressive prefix for audio-token generation, providing a
pre-synthesis interface for layout-level control.

\section{Experiments}

We evaluate generation, structural control, layout manipulation, ablations,
and subjective quality against representative text-to-music systems.
For our framework, we consider two layout conditions. In the reference-layout
condition, the model receives a MusicLayout derived from the target music. This
condition jointly tests whether MusicLayout captures the target's musical
organization and whether the model can use it to control the structure of the
generated audio. In the generated-layout condition, the model predicts a layout
from the text prompt before generating audio, evaluating the complete
end-to-end text-to-music process.

\subsection{Experimental Setup}

\paragraph{Datasets.}
We used three datasets with complementary roles.  FreeMIDI~\cite{freemidi}
provided the training data and a disjoint in-domain evaluation set.
MidiCaps~\cite{Melechovsky2024midicaps} was used to evaluate generalization to
out-of-domain MIDI, while MuChin~\cite{wang2024muchin} provided an out-of-domain
evaluation on real audio.  Their details are summarized in
Table~\ref{tab:datasets}.

\begin{table}[t]
\centering
\footnotesize
\setlength{\tabcolsep}{2.8pt}
\begin{tabular}{@{}llrr@{}}
\toprule
Dataset & Audio source & Train & Eval. \\
\midrule
FreeMIDI & MIDI-synth. & 24,474 & 2,719 \\
MidiCaps & MIDI-synth. & -- & 1,040 \\
MuChin & Real, separated accomp. & -- & 1,000 \\
\bottomrule
\end{tabular}
\caption{Datasets used in our experiments.}
\label{tab:datasets}
\end{table}

\textbf{FreeMIDI.}
We collected a subset of FreeMIDI and retained only pieces longer than 15 seconds.
For each remaining MIDI file, we extracted a MusicLayout and synthesized its
audio, forming a layout--audio pair. We then
provided the audio to MOSS-Music~\cite{mossmusic2026}, an audio-understanding language model, to produce
the generation prompt, yielding aligned prompt--layout--audio triples.

\textbf{MidiCaps.}
MidiCaps already provides natural-language captions.  We selected a subset with
a balanced distribution of style labels and removed MIDI files that overlap
with FreeMIDI.  For each remaining piece, we extracted a layout and synthesized
audio from the MIDI, directly pairing both with its caption to form evaluation
triples.

\textbf{MuChin.}
MuChin contains vocal music with Chinese captions but no MIDI, and thus cannot
support the reference-layout condition.  We used DeepSeek-V4-Flash~\cite{xu2026deepseek} to translate the captions into
English and remove vocal-related descriptions.  We further separated the
accompaniment from the vocals using a community-trained BS-RoFormer
model~\cite{lu2024music} and
used only the separated accompaniment for metric computation.  Generation
conditions that require an externally supplied layout---the
reference-layout and shuffled-layout-inference conditions---are not reported
on MuChin.

\paragraph{Compared systems.}
We compared against MusicGen \cite{copet2024musicgen}, ACE-Step~1.5 and
ACE-Step XL-Turbo \cite{gong2026ace}, and Stable Audio~3 Medium
\cite{evans2026stable}.  We evaluated our model under either the
generated-layout or the reference-layout condition.  To control for the effect of additional
training data, we also finetuned ACE-Step~1.5 on the same data using the same LM
adaptation setup as our model. We denote this matched-data no-layout control as
ACE-Step~1.5-FT. It directly predicts audio from the conditioning context $c$
without MusicLayout tokens, serving as both the primary fair comparison for explicit layout planning and
the first step of the progressive ablation. Systems shared evaluation items,
prompts, and seeds when supported, and baselines used officially recommended
decoding settings. Systems without a layout representation
targeted each item's reference duration.
For long-form MusicGen generation, we followed its
official continuation mechanism.
For our model, the generation
duration is instead determined by the end time of the conditioning layout. In
the generated-layout condition, this duration is therefore predicted by the
model rather than specified externally, whereas the reference-layout
condition matches the reference duration because its layout is derived from
the target music.

\paragraph{Evaluation protocol.}

We report Fr\'echet Audio Distance (FAD)~\cite{roblek2019FAD} for acoustic
distribution similarity, PaSST-KL~\cite{koutini2022efficient} for sound-event
agreement, SSIM~\cite{wang2004image} for local time--frequency similarity,
and CLAPScore~\cite{clap} for text--audio alignment.

For long-range structure, SCM Energy Distance
\cite{de2022measuring,szekely2013energy} compares corpus-level structural
complexity distributions, while $F_{0.5}$ and $F_{3.0}$ acoustic-boundary
agreement~\cite{turnbull2007supervised} measure fine- and coarse-grained
transition alignment within each piece.
To examine how layout manipulation changes long-range organization, we also
visualize recurrence-based self-similarity matrices (SSMs), following the
structural visualization used in Stable Audio~2 and its underlying music
structure analysis method~\cite{evans2024longform,serra2014unsupervised}.

\paragraph{Implementation details.}

We finetuned ACE-Step~1.5's 1.7B-parameter LM using six A800 GPUs, a
4,096-token context, and an effective batch size of 48. We selected the
checkpoint with the lowest development layout-to-audio loss. Layout and audio
decoding used
temperature/top-$p$ values of $0.8/0.95$ and $0.9/0.95$.

\subsection{Results}

\subsubsection{Overall Generation Performance}

\begin{table*}[t]
\centering
\small
\setlength{\tabcolsep}{2.2pt}
\begin{tabular}{@{}lllrrrrrrr@{}}
\toprule
Dataset & Group & System & FAD$\downarrow$ & KL$\downarrow$ & SSIM$\uparrow$ &
CLAP$\uparrow$ & SCM$\downarrow$ & $F_{0.5}$$\uparrow$ & $F_{3.0}$$\uparrow$ \\
\midrule
FreeMIDI & Baselines & ACE-Step 1.5
& 2.874 & 0.789 & 0.159 & 0.284 & 0.117 & 0.637 & 0.831 \\
& & ACE-Step XL-Turbo
& \underline{2.584} & 0.734 & 0.181 & \underline{0.300} & 0.063 & 0.619 & 0.811 \\
& & MusicGen-Large
& 3.706 & 0.834 & \underline{0.189} & 0.280 & 0.375 & 0.558 & 0.757 \\
& & Stable Audio 3 Medium
& 3.014 & 0.838 & \textbf{0.213} & \textbf{0.378} & 0.181 & 0.616 & 0.809 \\
\cmidrule(lr){2-10}
& Our model & Reference layout
& 2.610 & 0.740 & 0.183 & 0.238 & \underline{0.054} & \textbf{0.645} & \textbf{0.838} \\
& & Generated layout
& \textbf{2.495} & \textbf{0.719} & 0.153 & 0.241 & 0.206 & \underline{0.643} & \underline{0.834} \\
\cmidrule(lr){2-10}
& Matched-data & ACE-Step 1.5-FT (no layout)
& 3.345 & 0.956 & 0.154 & 0.219 & 0.380 & 0.577 & 0.739 \\
& controls & Shuffled-layout training
& 2.756 & \underline{0.721} & 0.153 & 0.216 & 0.100 & 0.607 & 0.810 \\
& & Shuffled-layout inference
& 2.636 & 0.779 & 0.176 & 0.230 & \textbf{0.038} & 0.642 & 0.831 \\
\midrule
MidiCaps & Baselines & ACE-Step 1.5
& 2.526 & 0.797 & 0.152 & 0.290 & 0.110 & \underline{0.632} & 0.828 \\
& & ACE-Step XL-Turbo
& 3.036 & 0.756 & 0.174 & \underline{0.313} & \underline{0.100} & 0.606 & 0.801 \\
& & MusicGen-Large
& 2.488 & 0.825 & \underline{0.180} & 0.258 & 0.379 & 0.545 & 0.746 \\
& & Stable Audio 3 Medium
& \textbf{2.123} & 0.726 & \textbf{0.229} & \textbf{0.352} & 0.315 & 0.616 & 0.812 \\
\cmidrule(lr){2-10}
& Our model & Reference layout
& \underline{2.149} & \underline{0.712} & 0.169 & 0.305 & \textbf{0.087} & \textbf{0.635} & \textbf{0.836} \\
& & Generated layout
& 2.303 & \textbf{0.699} & 0.140 & 0.287 & 0.804 & 0.630 & 0.830 \\
\cmidrule(lr){2-10}
& Matched-data & ACE-Step 1.5-FT (no layout)
& 2.740 & 0.877 & 0.153 & 0.234 & 0.351 & 0.560 & 0.736 \\
& controls & Shuffled-layout training
& 2.772 & 0.734 & 0.144 & 0.267 & 0.136 & 0.599 & 0.807 \\
& & Shuffled-layout inference
& 2.192 & 0.740 & 0.165 & 0.293 & 0.117 & \textbf{0.635} & \underline{0.834} \\
\midrule
MuChin & Baselines & ACE-Step 1.5
& 2.480 & 0.685 & 0.129 & 0.289 & \underline{0.186} & 0.588 & 0.797 \\
& & ACE-Step XL-Turbo
& \textbf{1.994} & \textbf{0.617} & 0.147 & \underline{0.317} & 0.611 & 0.566 & 0.779 \\
& & MusicGen-Large
& 3.279 & 0.717 & \underline{0.156} & 0.241 & 0.215 & 0.517 & 0.756 \\
& & Stable Audio 3 Medium
& \underline{2.007} & \underline{0.662} & \textbf{0.177} & \textbf{0.365} & \textbf{0.132} & 0.573 & 0.794 \\
\cmidrule(lr){2-10}
& Our model & Generated layout
& 3.456 & 0.671 & 0.116 & 0.285 & 0.655 & \underline{0.594} & \underline{0.809} \\
\cmidrule(lr){2-10}
& Matched-data & ACE-Step 1.5-FT (no layout)
& 3.124 & 0.844 & 0.123 & 0.222 & 1.458 & 0.535 & 0.726 \\
& controls & Shuffled-layout training
& 3.714 & 0.670 & 0.120 & 0.286 & 0.626 & \textbf{0.600} & \textbf{0.820} \\
\bottomrule
\end{tabular}
\caption{Objective results. CLAP, KL, and SCM abbreviate CLAPScore, PaSST-KL,
and SCM Energy Distance. Bold and underline mark the first- and second-ranked
numerical values per dataset, with rounded ties bold. The three matched-data
controls remove MusicLayout or mismatch it in training or inference.}
\label{tab:main_results}
\end{table*}

Table~\ref{tab:main_results} summarizes the objective results. No single
system dominates all aspects of generation.
Our model with generated layouts records the lowest FAD and PaSST-KL on
FreeMIDI. On MidiCaps, it also records the lowest PaSST-KL, suggesting that
layout-conditioned generation can remain close to the reference acoustic-event
distribution despite out-of-domain MIDI captions. Its lower CLAPScore relative
to Stable Audio and ACE-Step XL-Turbo, however, shows that explicit planning
may not uniformly improve global text--audio correspondence.

The matched-data no-layout control provides a more direct test of the
contribution of explicit layout planning. It is weaker than the original
ACE-Step~1.5 on nearly all metrics. This degradation
likely reflects the training-data difference: the original model was trained
on high-quality real recordings, whereas our finetuning data consists of
MIDI-synthesized audio with lower audio quality and timbral fidelity, as well
as a different acoustic distribution. Since our model and this control share
the same finetuning data and LM adaptation setup, their comparison isolates
the effect of explicit layout planning. Relative to this control, the
reference-layout condition has more favorable values on all seven metrics for
FreeMIDI and MidiCaps. The generated-layout condition does so on six metrics
for FreeMIDI and five for MidiCaps and MuChin.
These
results show that explicit layout planning substantially improves generation under matched-data
conditions and can partly offset the limitations of weaker training audio.

Overall, our model is competitive on the two MIDI-derived datasets,
indicating that explicit layout planning contributes to overall music
generation performance.  Its FAD, PaSST-KL, and SSIM results on MuChin are
less favorable, however.  This difference likely arises from an acoustic-domain
shift: our model is trained on MIDI-synthesized instrumental audio, whereas
MuChin consists of accompaniments separated from real recordings.

\subsubsection{Structural Control}

The reference-layout condition records the highest boundary scores shown for
FreeMIDI, the highest MidiCaps $F_{3.0}$, and a tie for the highest MidiCaps
$F_{0.5}$.
Because the reference layouts are derived from the target music, these results
provide evidence that MusicLayout captures target organization and that
conditioning on it can guide corresponding structure in the generated audio.

With model-generated layouts, our model also shows strong structural
organization without access to the target-derived layout used in the
reference-layout condition. On FreeMIDI, both boundary scores are numerically
above all baselines. On MidiCaps, $F_{3.0}$ is numerically above all baselines,
while $F_{0.5}$ remains competitive. On MuChin, both boundary scores are
numerically above all baselines, showing that this advantage persists under an
acoustic-domain shift from MIDI-synthesized training data to accompaniments
separated from real recordings. These results
indicate that explicit intermediate planning improves the organization of
long-range musical structure in end-to-end generation.

The comparison with the matched-data no-layout control further isolates this
structural benefit. Both layout conditions have higher boundary scores than
the no-layout control wherever available, and the generated-layout condition
retains this advantage across all three datasets.
Thus, under the same MIDI-synthesized training data, explicit layout planning
improves boundary agreement in both controlled and automatic generation.

SCM Energy Distance measures agreement between dataset-level structural
complexity distributions rather than song-level layout correctness.  On the
two MIDI-derived evaluation sets, our model in the reference-layout condition
records the lowest observed SCM Energy Distance among the non-ablation systems
, indicating that the specified layouts are realized with corpus-level
structural complexity close to the references.
On MuChin,
the SCM Energy
Distance in the generated-layout condition is comparatively high. This weaker agreement is consistent with the
training--evaluation
shift from MIDI-synthesized instrumental music to accompaniments separated
from real vocal recordings, which have substantially different
structural-complexity distributions. However, with training data controlled,
the generated-layout condition has lower SCM than the no-layout condition.
This comparison suggests a structural benefit from layout planning.

\begin{figure}[t]
\centering
\includegraphics[width=0.86\linewidth]{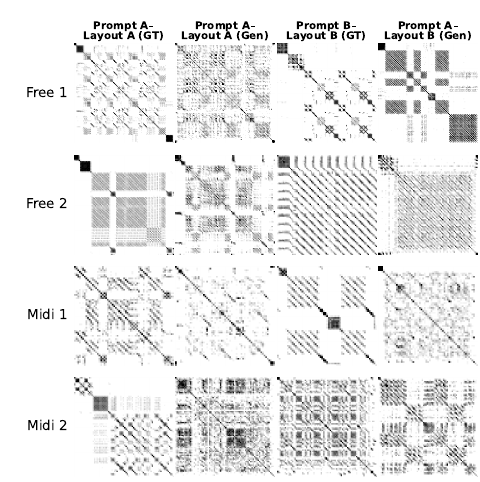}
\caption{Structural effects of layout manipulation for two FreeMIDI and two
MidiCaps examples. Each row presents, from left to right, the target GT under
Prompt A and Layout A, generation under the same prompt and reference layout,
the layout-donor GT under Prompt B and Layout B,
and generation from Prompt A paired with Layout B. For clear visualization,
SSM intensities are normalized independently within each panel. Darker regions
within a panel indicate stronger recurrence. Audio signals in each row are
cropped to their common minimum duration.}
\label{fig:layout_manipulation_ssm}
\end{figure}

\subsubsection{Structural Effects of Layout Manipulation}

Figure~\ref{fig:layout_manipulation_ssm} presents four separate
layout-manipulation cases, with two drawn from FreeMIDI and two from MidiCaps.
These cases examine how layout manipulation changes the generated structure
while the text prompt is held fixed. In each
row, item A serves as the target and provides Prompt A and its reference Layout
A. A different item B supplies Layout B, which we refer to as the donor layout.
The SSM for item B is included to visualize the recurrence structure associated
with Layout B. We refer to generation from Prompt A and Layout A as the matched
generation, and generation from the same prompt and Layout B as the manipulated
generation. In these examples, the matched generation
shows recurrence patterns similar to the target reference, while the
manipulated generation loses some target-aligned patterns and exhibits patterns
closer to the donor reference. Because the manipulated generation combines the
target prompt with the donor layout, it can retain structural characteristics
associated with the target while adopting aspects of the donor's organization.
Together, these cases illustrate that MusicLayout can be manipulated before
audio generation to control the structure of the generated music.

\subsubsection{Fine-Grained Regional Controls}

\begin{figure}[t]
\centering
\includegraphics[width=0.94\textwidth]{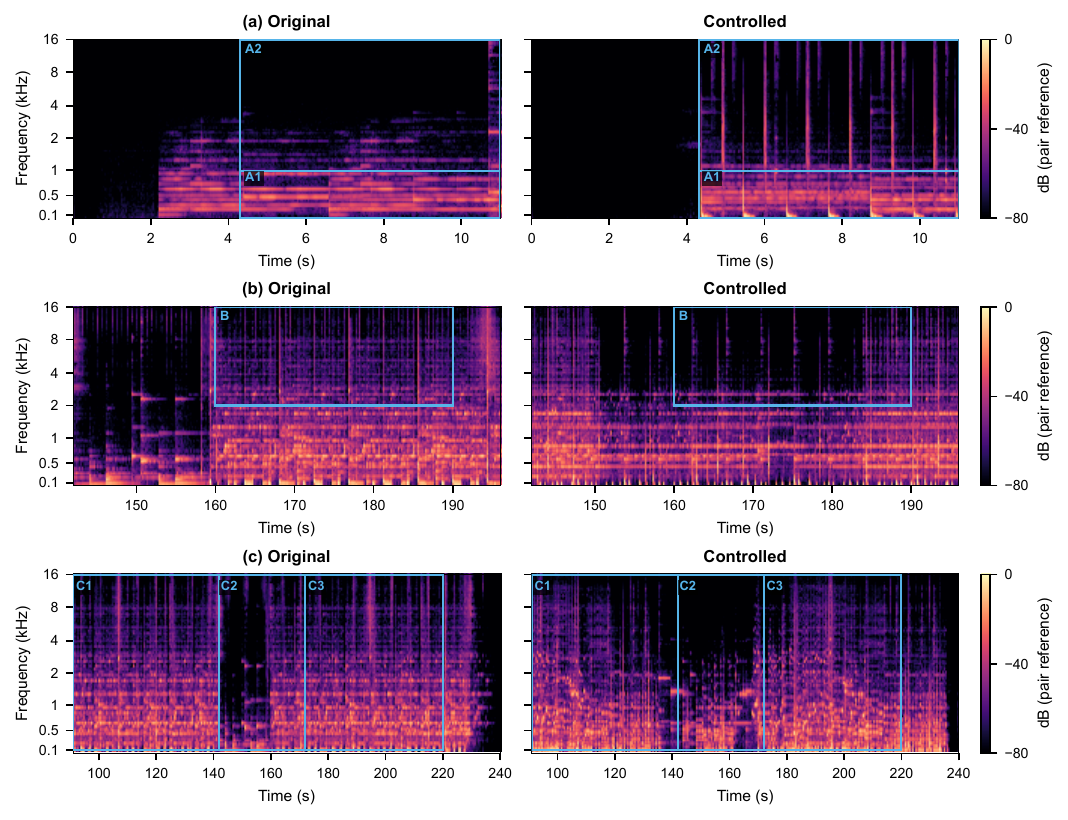}
\captionof{figure}{Paired log-mel spectrograms for three fine-grained MusicLayout
controls: (a) role replacement over 0--11 seconds, (b) reduced density and
energy over 142--196 seconds, and (c) a form rewrite over 91--240 seconds.
Light-blue boxes use matched time--frequency coordinates: A1/A2 mark
low-frequency and broadband changes, B marks reduced broadband activity, and
C1--C3 mark successive stages of the form rewrite. Each pair shares its 0-dB
reference and 80-dB range. Intensities are comparable within a row. Time axes
retain full-song coordinates.}
\label{fig:fine-grained-control-spectrograms}
\end{figure}

We further tested three direct edits to a single 240-second MusicLayout,
holding the text prompt and sampling seeds fixed so that each controlled
condition differed from the original only in the declared MusicLayout fields.
Figure~\ref{fig:fine-grained-control-spectrograms} compares the original and
controlled audio over the affected interval.

First, in the 0--11-second introduction, we replaced sparse, high-energy
strings and synthesizer strings with dense, high-energy drums and sparse,
high-energy bass. This edit was expected to add stronger low-frequency rhythmic
content and broadband percussive transients. The controlled spectrogram shows
both effects: energy below 250~Hz becomes dominant and repeated vertical
transients appear across the band.

Second, over 142--196 seconds, we changed every active role to sparse density
and medium energy while leaving the segment boundaries, labels, families, and
instrument identities unchanged. The expected result was a less dense and
less intense realization of the same region. Relative to the original, the
controlled excerpt has visibly reduced broadband activity and its RMS level
decreases by 15\% (1.4~dB).

Third, we preserved the complete 0--91-second prefix and rewrote the remaining
form from a largely repeated chorus sequence into a breakdown, bridge, build,
solo, chorus, and outro, together with their associated roles and recurrence
families. The expected result was a new long-range progression after the
preserved prefix. The controlled spectrogram exhibits the intended succession
of contrasting regimes, including the reduced texture of the breakdown and
bridge, the subsequent build, and the denser solo and chorus.
Thus, all three edits produce changes consistent with their specified local or
regional controls.

\subsubsection{Ablation Study}

The three matched-data controls in Table~\ref{tab:main_results} form a
progressive ablation of MusicLayout. First, the finetuned ACE-Step~1.5 control
removes MusicLayout entirely. Second, shuffled-layout training introduces
MusicLayout but pairs each training item with a layout from another item, so
the model learns from mismatched layout--audio pairs. At inference, this model
generates its own layout from the prompt before audio generation, following the
same end-to-end procedure as the full generated-layout condition. Third, shuffled-layout
inference restores correctly paired layouts during training but replaces the
conditioning layout with one from another item at inference.

Shuffled-layout training has favorable values on five of seven metrics relative to the no-layout control for
each of the three datasets. This pattern suggests that
layout-shaped supervision may retain
a generic structural benefit even when item-level information is incorrect.
On the two MIDI-synthesized datasets, our full model with generated layouts
has favorable values on five of seven metrics relative to shuffled-layout training for both
FreeMIDI and MidiCaps. Because both conditions generate layouts at inference,
this pattern suggests that correct layout--audio correspondence during training
provides information beyond the presence of layout-shaped tokens and can
improve long-form music generation when the evaluation domain matches the type
of audio used for training. On MuChin, the two conditions obtain broadly similar
results. This may result from the acoustic-domain shift discussed above, since our
model was trained on MIDI-synthesized audio rather than accompaniments from
real recordings.

Holding correctly paired training fixed, the reference-layout condition has favorable
FAD, KL, SSIM, CLAPScore, and $F_{3.0}$ values relative to shuffled-layout inference on both MIDI-derived
datasets. This indicates that item-matched layouts provide a more informative
inference-time conditioning signal than a plausible but unrelated layout.

A shuffled layout remains musically plausible because it comes from a real
piece and retains valid sections, timing, repetitions, and arrangements.
Shuffling can therefore preserve the corpus-level layout distribution while
breaking item-level correspondence, so low SCM alone does not demonstrate
correct layout control.

\subsection{Subjective Evaluation}

\begin{table}[t]
\centering
\footnotesize
\setlength{\tabcolsep}{2.8pt}
\begin{tabular}{@{}lrrrrrr@{}}
\toprule
& \multicolumn{3}{c}{All ($N=59$)}
& \multicolumn{3}{c}{Exp. ($N=36$)} \\
\cmidrule(lr){2-4}\cmidrule(lr){5-7}
System & F & M & T & F & M & T \\
\midrule
GT
& 3.200 & 3.490 & 3.408
& 3.000 & 3.361 & 3.300 \\
ACE~1.5
& 3.283 & 3.329 & 3.225
& 3.031 & 3.342 & 3.108 \\
ACE-Step~1.5-FT
& 2.931 & 2.632 & 2.702
& 2.631 & 2.386 & 2.661 \\
Ours-Ref
& 3.064 & 3.012 & 3.093
& 2.989 & 2.917 & 3.183 \\
Ours-Gen
& 2.914 & 2.586 & 2.869
& 2.733 & 2.783 & 2.839 \\
\bottomrule
\end{tabular}
\captionof{table}{Subjective ratings on FreeMIDI and MidiCaps (0--5). F/M/T
denote fidelity, musical impression, and text consistency. Exp.\ denotes
experienced listeners, and FT denotes matched-data finetuning.}
\label{tab:subjective-main}

\vspace{4pt}
\setlength{\tabcolsep}{2.5pt}
\begin{tabular}{@{}lr@{}}
\toprule
System & Score \\
\midrule
ACE-Step~1.5 & 2.319 \\
ACE-Step~1.5-FT & 1.444 \\
Ours (reference layout) & 2.556 \\
Ours (generated layout) & 2.044 \\
\bottomrule
\end{tabular}
\captionof{table}{GT-reference structural similarity rated by
music-experienced participants on the same MIDI-derived items (0--5).}
\label{tab:subjective-structure}
\end{table}

To enable a direct comparison with the reference-layout condition, we excluded
MuChin because it does not provide reference layouts. We sampled 10 items each
from FreeMIDI and MidiCaps and assigned them repeatedly to 27 participants,
including 15 with prior music experience. Participants rated audio fidelity,
musical impression, and text consistency on a 0--5 scale, with one decimal
place allowed. Each evaluation compared all five systems,
yielding 59 ratings per system, including 36 ratings from experienced
listeners.

Table~\ref{tab:subjective-main} shows that comparisons with the matched-data
no-layout control vary with layout condition and listener experience. Across
all participants, the reference-layout condition has higher observed scores on
all three dimensions. Generated-layout scores are broadly comparable,
with similar fidelity and musical impression but higher text consistency.
Within the experienced subset, both layout conditions score numerically higher
than the control across all dimensions. Notably, the reference-layout condition
also has higher observed text consistency than the original ACE-Step~1.5 among
experienced listeners (3.183 vs.\ 3.108), suggesting that an explicit
MusicLayout can provide an additional control signal for steering music
generation. This pattern may arise because
music-experienced listeners are more attentive to structural organization when
evaluating generated music. ACE-Step~1.5 has the highest observed fidelity
score across the five systems (3.283), above GT (3.200). As discussed above,
this difference is consistent with the acoustic-quality gap between the
MIDI-synthesized data in our experiments and the real recordings used to train
ACE-Step~1.5. It
underscores the need for matched-data controls to
isolate layout planning from training-data differences.

Using the same dataset scope, we conducted a follow-up on a subset of the
previously sampled items. A subset of the experienced participants was shown
the GT reference for each item and asked to rate the structural
similarity of the remaining outputs to the references. The reference-layout
condition received a higher observed structural
similarity than ACE-Step~1.5, providing subjective evidence that MusicLayout
captures reference organization and carries it into generated audio. The
generated-layout condition was slightly weaker because coarse dataset prompts
do not uniquely specify the GT layout. Once predicted, the layout steers
generation toward its own plausible structure. The objective structural
metrics above suggest that these outputs can remain structurally organized even
when they do not reproduce the particular GT structure.

\FloatBarrier
\section{Conclusion}

We presented MusicLayout, an explicit, time-aligned representation of section
organization, development, and instrument arrangement for a unified
autoregressive audio LM. Generated before audio tokens, it makes structural
planning inspectable and adjustable. The same LM plans the layout and predicts
audio tokens conditioned on it while the synthesis components remain frozen.
Our results suggest that MusicLayout provides an interpretable interface that
supports long-range structural organization and layout-level control.

Despite these benefits, MusicLayout adjusts structural plans only before
synthesis and cannot edit existing audio or regenerate selected regions.
Our reliance on MIDI-synthesized training audio may also limit audio fidelity.
Future work will pursue finer control and higher-quality, structurally aligned
audio data.

\section*{Acknowledgments}

This work was supported by Ant Group.

\appendix

\section{MusicLayout Representation and Annotation}

\subsection{Serialization Grammar}

\MusicLayout{} is serialized as a discrete text sequence with one optional
piece-level family block and one required segment block. A family records a
piece-local identifier, its member segments, its cardinality, and whether the
members are adjacent. Each segment records an integer-second time span, a
functional section label, a texture, an arrangement-change label, its family
membership and family role, the degree of variation from the family prototype,
a repetition flag, and a list of active instrument tuples. The grammar used in
all reported experiments is summarized below:

\begin{quote}
\scriptsize\ttfamily
<layout>\\
<families>\\
<fam> id=fam\_K members=sI,sJ count=N adj=true|false </fam>\\
</families>\\
<segments>\\
<seg> id=sI time=A-B label=L texture=T change=C\\
\hspace*{1em}fam=fam\_K fam\_role=R variation=V repeat=true|false\\
<roles> instrument:register:density:energy | ... </roles>\\
</seg>\\
</segments>\\
</layout>
\end{quote}

Family identifiers are local to a piece rather than global semantic labels.
For example, \texttt{fam\_0} denotes one recurring material within the current
piece and may refer to unrelated material in another piece. The extraction
pipeline retains singleton material as an explicitly declared one-member
family with \texttt{count=1} and \texttt{adj=true}. Its segment uses
\texttt{variation=unique} and \texttt{repeat=false}. For a recurring family,
the first chronological member serves as its variation prototype and is marked
\texttt{same}. Later members are marked \texttt{same}, \texttt{light\_var}, or
\texttt{strong\_var} according to their distance from that prototype.

\subsection{Closed Vocabularies}

Table~\ref{tab:supp-structural-vocabulary} gives the structural vocabularies.
The \texttt{change} field describes the segment's arrangement-level relation
to its predecessor, whereas \texttt{fam\_role} describes its role in the
piece-level recurrence organization. These fields are distinct from the
functional section label.

\begin{table*}[t]
\centering
\small
\setlength{\tabcolsep}{5pt}
\begin{tabular}{@{}p{0.17\textwidth}p{0.76\textwidth}@{}}
\toprule
Field & Allowed values \\
\midrule
Section label & \texttt{intro}, \texttt{verse}, \texttt{prechorus},
\texttt{chorus}, \texttt{bridge}, \texttt{breakdown}, \texttt{outro},
\texttt{transition}, \texttt{hook}, \texttt{solo}, \texttt{build} \\
Texture & \texttt{layered}, \texttt{rhythm\_driven},
\texttt{melodic\_front}, \texttt{harmonic\_bed}, \texttt{percussive},
\texttt{build\_up}, \texttt{sparse\_pulse}, \texttt{lead\_front},
\texttt{contrast}, \texttt{other} \\
Change & \texttt{entry}, \texttt{continuation}, \texttt{lift},
\texttt{drop}, \texttt{contrast}, \texttt{outro} \\
Family role & \texttt{intro\_anchor}, \texttt{primary\_repeat},
\texttt{secondary\_repeat}, \texttt{adjacent\_variant},
\texttt{outro\_anchor}, \texttt{unique} \\
Variation & \texttt{unique}, \texttt{same}, \texttt{light\_var},
\texttt{strong\_var} \\
Register & \texttt{low}, \texttt{mid}, \texttt{high} \\
Density & \texttt{sparse}, \texttt{med}, \texttt{dense} \\
Energy & \texttt{low}, \texttt{med}, \texttt{high} \\
\bottomrule
\end{tabular}
\caption{Closed structural and arrangement vocabularies in MusicLayout.}
\label{tab:supp-structural-vocabulary}
\end{table*}

\begingroup
\sloppy
The instrument field uses 25 compact categories:
\path{acoustic_piano}, \path{electric_piano},
\path{plucked_keyboard}, \path{mallet_bell}, \path{organ},
\path{accordion_harmonica}, \path{acoustic_guitar},
\path{electric_guitar}, \path{bass}, \path{strings},
\path{orchestral}, \path{synth_strings}, \path{choir_voice},
\path{brass}, \path{sax}, \path{woodwind},
\path{synth_lead}, \path{synth_pad}, \path{synth_fx},
\path{world_plucked}, \path{percussion},
\path{reverse_cymbal}, \path{sound_fx}, \path{drums}, and
\path{other}. General MIDI program numbers are mapped deterministically to
these categories. Instrument entries retain the segment-local register,
density, and energy attributes. These attributes are defined at the instrument
tuple level rather than duplicated as separate segment-level fields.
\par
\endgroup

\subsection{Variation Strength}

For each recurring family, we compare every later segment with the first family
member. Let $d_r$ be the Jaccard distance between active-role sets, and let
$d_e$, $d_d$, and $d_a$ be normalized differences in energy, note density, and
active-role count. The base distance is
\begin{equation}
d = 0.45d_r + 0.20d_e + 0.20d_d + 0.15d_a.
\end{equation}
The implementation adds small deterministic adjustments when coarse energy,
density, texture, arrangement-change, repeat-neighbor, or source-relation
evidence differs from the family prototype. We serialize scores at most $0.15$
as \texttt{same}, scores in $(0.15,0.40]$ as \texttt{light\_var}, and larger
scores as \texttt{strong\_var}. This field
therefore records an interpretable, rule-derived degree of variation rather
than a human rating.

\subsection{Validation}

Before a generated layout can condition audio decoding, the parser checks the
outer layout and segment wrappers, unique segment identifiers, nondecreasing
integer-second start times, positive spans, valid closed-vocabulary values,
four-field instrument tuples, family-member counts, and references from
segments to declared families. It also rejects legacy bar spans, legacy role
tags, segment-level energy/density fields, internal MIDI program slugs, and
audio-code tokens inside the layout. A failed layout is not silently repaired
or rewritten.

\subsection{Annotation Pipeline}

The annotation pipeline begins from aligned MIDI and rendered WAV files. It
extracts bar-level activity and instrument statistics, combines change and
repetition evidence into time-aligned segments, groups structurally related
segments into piece-local material families, and derives the section, texture,
change, family, variation, and instrument attributes serialized above. MIDI
programs are used only as annotation evidence. The learning target remains
audio-token generation. Segment spans are converted to audio time and rounded
to integer seconds, with a minimum duration of one second after rounding.

The extraction procedure is deterministic once its feature and boundary
settings are fixed. Algorithm~\ref{alg:musiclayout-extraction} summarizes its
execution order.
The symbolic representation supplies the structural evidence. The paired
waveform is used to place the resulting spans on the rendered-audio timeline.

\begin{algorithm}[!tbp]
\caption{MusicLayout extraction from aligned symbolic music and audio}
\label{alg:musiclayout-extraction}
\footnotesize
\setlength{\baselineskip}{8.8pt}
\begin{algorithmic}[1]
\Require MIDI performance $M$, aligned waveform duration $D$, and fixed extraction settings $\Theta$
\Ensure Valid serialized layout $L$, or rejection symbol $\bot$
\Function{ExtractMusicLayout}{$M,D,\Theta$}
  \State $(N,P,T,B,E) \gets \Call{ParseSymbolic}{M}$ \Comment{notes, programs, MIDI duration, bars, beats}
  \For{$b \in B$}
    \State $f_b \gets \Call{AggregateBar}{N,P,b}$ \Comment{density, energy, pitch, and six broad-role activities}
  \EndFor
  \State $q \gets \Call{AdjacentChangeScores}{f}$
  \State $R \gets \Call{BinarySegmentationCandidates}{f,D}$
  \State $q \gets \Call{AddBoundarySupport}{q,R,f}$
  \State $n \gets \min(14,\max(5,\Call{Round}{D/28}))$
  \State $C \gets \Call{SelectPeaks}{q,\operatorname{mean}(q)+0.34\operatorname{std}(q),3\ \mathrm{bars},n}$
  \State $C \gets \Call{RefineLongSpans}{C,f,B}$ \Comment{adjacent-repeat and local-change tests}
  \State $C \gets \Call{InsertMicroBoundaries}{C,f,B}$
  \State $S \gets \Call{ScaleBoundariesToAudio}{C,B,T,D}$
  \State $S \gets \Call{MergeShortSegments}{S,8\ \mathrm{s}}$
  \State $S \gets \Call{RefineOpeningWithBeats}{S,E,f,B,T,D}$
  \For{$s_i \in S$}
    \State $z_i \gets \Call{AggregateSegment}{\{f_b:b\subseteq s_i\}}$
    \State $e_i \gets \Call{AggregateInstrumentStatistics}{N,P,s_i}$
  \EndFor
  \State $\mathcal{F}\gets\emptyset$
  \For{$i=1,\ldots,|S|$} \Comment{chronological greedy centroid assignment}
    \State $k^* \gets \arg\max_k\operatorname{cos}(z_i,\mu_k)$
    \If{$\mathcal{F}=\emptyset$ or $\operatorname{cos}(z_i,\mu_{k^*})<0.965$}
      \State $\phi(i)\gets\Call{NewFamily}{\mathcal{F},z_i}$
    \Else
      \State $\phi(i)\gets k^*,\quad \mu_{k^*}\gets\Call{MemberMean}{\{z_j:\phi(j)=k^*\}}$
    \EndIf
  \EndFor
  \For{$i,j \in \{1,\ldots,|S|\}$}
    \State $A_{ij} \gets \operatorname{cos}(z_i,z_j)$ \Comment{relation evidence, not family assignment}
  \EndFor
  \State $(\ell_i,t_i,c_i,u_i)_{i=1}^{|S|} \gets \Call{AssignDescriptors}{S,z,A,\mathcal{F},\phi}$
  \For{$F_k \in \mathcal{F}$}
    \State $p_k \gets \min\{i:\phi(i)=k\}$
    \For{$i:\phi(i)=k$}
      \State $v_i \gets \Call{VariationClass}{s_i,s_{p_k},\Theta_{\mathrm{variation}}}$
    \EndFor
  \EndFor
  \State $K \gets \Call{MapProgramsToCategories}{P}$
  \For{$s_i \in S$}
    \State $r_i \gets \Call{QuantizeInstrumentTuples}{e_i,K}$
  \EndFor
  \State $L \gets \Call{SerializeAndRound}{\mathcal{F},\{s_i,\ell_i,t_i,c_i,u_i,v_i,r_i\}_{i=1}^{|S|}}$
  \If{$\neg\Call{Validate}{L}$}
    \State \Return $\bot$
  \EndIf
\State \Return $L$
\EndFunction
\end{algorithmic}
\end{algorithm}

\section{Dataset Construction and Provenance}

\subsection{FreeMIDI}

FreeMIDI~\cite{freemidi} supplies the training corpus and an in-domain evaluation set. We
retained pieces longer than 15 seconds, extracted one MusicLayout from each
retained MIDI file, and synthesized the aligned audio at 44.1~kHz using the
FluidSynth~\cite{fluidsynth} interface provided by
PrettyMIDI~\cite{raffel2014prettymidi}, with
\texttt{MuseScore\_General.sf2}~\cite{musescoregeneralsf2}. Of 27,237
extracted items, 27,229 received a full-song caption from
MOSS-Music-8B-Instruct~\cite{mossmusic2026}. Caption decoding was greedy. The prompt requested one
concise English paragraph grounded only in the audio, covering style, audible
instrumentation and roles, rhythm, texture, energy, musical development, and
mood. It explicitly prohibited the use of filenames, MIDI metadata, external
labels, bullet lists, and unsupported claims.

Requiring at least 15 seconds of audio and a nonempty supervised audio-code
span retained 27,198 items. Five additional items were removed because their
complete planning sequences exceeded the 4,096-token context. The resulting
27,193 items were split deterministically into 24,474 training and 2,719
development items with seed 20260501. The development split is also the
in-domain evaluation set.

\subsection{MidiCaps}

MidiCaps~\cite{Melechovsky2024midicaps} is an out-of-domain MIDI evaluation
set. We first sampled 1,100 items
with seed 42 after requiring a duration of at least 120 seconds, at least two
instruments, and a nonempty caption. Genre-quota sampling followed by
instrument-diverse greedy selection covered 40 source genre labels. Because
the labels are multi-valued, their frequencies need not sum to the number of
items. The most frequent labels in the final set are electronic (581), pop
(463), rock (208), classical (108), soundtrack (102), ambient (81), and jazz
(59). Exact SHA-256 comparison of the MIDI files against the complete FreeMIDI
corpus removed 60 overlaps, leaving 1,040 unique items. The source location and
MIDI SHA-256 jointly define item identity. We then extracted MusicLayouts and
synthesized aligned audio using the same renderer as for FreeMIDI. Original
MidiCaps captions were retained as prompts, and no MidiCaps item was used for
training.

\subsection{MuChin}

MuChin~\cite{wang2024muchin} provides an out-of-domain real-audio evaluation. Because it does not
provide MIDI, it cannot supply a reference MusicLayout. We constructed a fixed
1,000-item manifest in the source metadata order. For each item,
DeepSeek-V4-Flash~\cite{xu2026deepseek} received the original full-song
description as context and a
set of non-vocal tags covering instrumentation, arrangement, style, tempo,
rhythm, and mood. The instruction required one concise English paragraph using
only these safe musical attributes and explicitly prohibited any mention or
implication of singers, vocals, lyrics, rap, choir, humming, or spoken dialogue.
Decoding used temperature 0.2 and top-$p$ 0.9. We appended the same explicit
instrumental-only control phrase to every final prompt.

We separated accompaniment from vocals with the Python toolkit
\texttt{audio-separator} v0.44.3~\cite{beveridge2026audioseparator} and the
BS-RoFormer~\cite{lu2024music} Viperx-1297 checkpoint~\cite{trvlvrbsroformer317},
distributed as
\path{model_bs_roformer_ep_317_sdr_12.9755.ckpt}.
We used the resulting accompaniment as the metric reference.
Reference-layout and inference-time shuffled-reference-layout conditions are
consequently not applicable to MuChin.

\begin{center}
\centering
\small
\setlength{\tabcolsep}{3pt}
\begin{tabular}{@{}lrrr@{}}
\toprule
Dataset & Train & Evaluation & Ref. layout \\
\midrule
FreeMIDI & 24,474 & 2,719 & Yes \\
MidiCaps & 0 & 1,040 & Yes \\
MuChin & 0 & 1,000 & No \\
\bottomrule
\end{tabular}
\captionof{table}{Dataset roles and final item counts.}
\label{tab:supp-datasets}
\end{center}

\section{Model Adaptation and Training}

\subsection{Training Sequences and Loss Masks}

The adapted model uses the 1.7B-parameter LM from
ACE-Step 1.5~\cite{gong2026ace}. Layout planning
and audio-token prediction share the same LM but use separate supervised target
spans. In the planning task, the prompt and chat context are inputs and the
MusicLayout tokens are targets. In the layout-to-audio task, the prompt and
ground-truth MusicLayout form the input prefix and only the following audio-code
tokens are targets. Chain-of-thought metadata is retained in the serialized
prefix but receives zero loss weight. The pretrained 5-Hz audio tokenizer and
DiT renderer~\cite{peebles2023dit} remain frozen in all stages.

Stage 1 trains only the newly introduced vocabulary rows in the input
embedding and LM head. Old vocabulary rows are masked, and no transformer layer
or final normalization parameter is updated. Stage 2 updates the LM while
alternating the planning and layout-to-audio tasks. The development
layout-to-audio loss determines checkpoint selection.

\begin{table*}[t]
\centering
\small
\setlength{\tabcolsep}{5pt}
\begin{tabular}{@{}p{0.22\textwidth}p{0.32\textwidth}p{0.36\textwidth}@{}}
\toprule
Setting & Stage 1 & Stage 2 \\
\midrule
Tasks & Layout planning & Layout planning and layout-to-audio \\
Epoch limit & 10 & 100 \\
Learning rate & $1.0\times10^{-4}$ & $3.0\times10^{-5}$ \\
Monitored development task & Planning & Layout-to-audio \\
Early-stopping patience & 5 & 10 \\
Trainable scope & New embedding/LM-head rows & LM \\
Loss-bearing targets & Layout span & Layout or audio span for the sampled task \\
\bottomrule
\end{tabular}
\caption{Stage-specific optimization protocol. The best Stage-2 checkpoint is
restored according to development layout-to-audio loss.}
\label{tab:supp-training-stages}
\end{table*}

Training used six 80-GB GPUs, a maximum sequence length of 4,096, per-rank
batch size 2, gradient accumulation 4, and effective batch size 48. Parameters
and AdamW~\cite{loshchilov2019adamw} optimizer states remained FP32, while
FSDP~\cite{zhao2023fsdp} computation used BF16.
Full-shard FSDP used original parameters, automatic wrapping, synchronized
module states, limited all-gathers, and replicated vocabulary parameters for
the Stage-1 row masks. The warmup ratio was 0.05 and the training random seed
was 42. The data split used seed 20260501.

\subsection{Matched-Data Controls}

The no-layout ACE-Step 1.5-FT control uses the same filtered training items,
audio codes, backbone LM, optimization precision, and adaptation budget as the
full model, but directly predicts audio tokens without a MusicLayout prefix.
The shuffled-layout-training control retains the MusicLayout-shaped prefix but
pairs each audio target with a layout from another item during training. At
inference it generates a layout from the prompt and then generates audio, as in
the full generated-layout condition. The shuffled-layout-inference control
uses the normally trained MusicLayout model but replaces the item-matched
reference layout with a valid layout from another item. Thus, the three controls
remove layout information, disrupt training-time correspondence, or disrupt
inference-time correspondence, respectively.

Both shuffled conditions use seed 20260711. For shuffled-layout training, the
24,474 training layouts form a one-to-one permutation of the same training
items, with zero fixed points. The 2,719 development items remain unchanged.
For shuffled-layout inference, FreeMIDI and MidiCaps are permuted independently
within their evaluation sets. This yields 3,759 unique target--donor pairs and
zero fixed points, preventing cross-dataset donors and self-matches. Donor
layouts are truncated or extended at segment boundaries to end at the target
item's requested duration.

\section{Baseline Selection and Exact Configurations}

\subsection{Selection Scope}

We restricted the baseline comparison to models with publicly released weights
and executable inference implementations. We further required compatibility
with instrumental prompt-to-music generation and several-minute output, either
through native variable-duration synthesis or through an established
continuation mechanism. These criteria allow the reported systems to consume
the same prompts and target durations and to be evaluated using the same audio
and structural metrics.

AudioLDM~2~\cite{liu2024audioldm2} and
Mustango~\cite{melechovsky2024mustango} were considered but not included in the
final comparison. Their released checkpoints and public inference protocols
are centered on short clips and do not provide a validated overlapping-context
continuation procedure for the several-minute setting used here. Concatenating
independently generated diffusion samples would introduce a different and
potentially discontinuous generation procedure. In contrast, the reported
diffusion baseline, Stable Audio 3 Medium~\cite{evans2026stable}, supports the required duration
natively.

\subsection{Executed Configurations}

Table~\ref{tab:supp-baseline-configs} reports the configurations used in the
final comparison. Each system targets the item's manifest duration.
ACE-Step-family and Stable Audio systems generate that duration natively,
without concatenating independently generated clips. The corresponding model
families are described by \citet{gong2026ace} and \citet{evans2026stable}.

\begin{table*}[t]
\centering
\small
\setlength{\tabcolsep}{4pt}
\begin{tabular}{@{}p{0.18\textwidth}p{0.29\textwidth}p{0.42\textwidth}@{}}
\toprule
System & Checkpoint/configuration & Executed inference settings \\
\midrule
MusicGen-Large & \texttt{facebook/musicgen-large} & 30-s window, 10-s audio context, 20-s retained hop, 51.2 tokens/s, 1,536 first-window and 1,024 continuation tokens, base seed 28602 \\
ACE-Step 1.5 & \texttt{acestep-v15-turbo}, \texttt{acestep-5Hz-lm-1.7B} & 8 inference steps, LM temperature/top-$p$ 0.9/0.95, native target duration, base seed 38602 \\
ACE-Step 1.5 XL-Turbo & \texttt{acestep-v15-xl-turbo}, \texttt{acestep-5Hz-lm-1.7B} & 8 inference steps, shift 3.0, LM temperature/top-$p$ 0.9/0.95, native target duration, base seed 38602 \\
Stable Audio 3 Medium & \texttt{stable-audio-3-medium} & 8 sampling steps, CFG 1.0, no negative prompt, FP16, chunked decoding, native target duration, base seed 48602 \\
\bottomrule
\end{tabular}
\caption{Exact configurations of the externally pretrained baselines. Base
seeds are combined with the global item index.}
\label{tab:supp-baseline-configs}
\end{table*}

Both ACE-Step configurations use the released LM-assisted inference path: the
LM first generates chain-of-thought metadata and audio semantic codes, which
are subsequently rendered by the DiT. We therefore treat ACE-Step 1.5 and
ACE-Step 1.5 XL-Turbo as hybrid LM--DiT baselines rather than pure diffusion
baselines. Stable Audio 3 Medium is the diffusion-only baseline in this
comparison.

For MusicGen-Large~\cite{copet2024musicgen}, the first window is generated from
the text prompt alone. Each later window receives the same text and the final
10 seconds of the accumulated waveform as audio context, generates at most
1,024 new codec tokens, and retains at most 20 seconds of new audio. The window
seed is the item seed plus the zero-based window index. The concatenated result
is cropped only at the final target duration.

\section{Inference and Evaluation Protocol}

\subsection{MusicLayout Inference}

Automatic inference first decodes the chain-of-thought metadata and
MusicLayout, validates the complete layout, and then continues with 5-Hz
audio-code tokens. Only the outer special-token boundaries are constrained.
the decoded metadata and layout body are not inserted, replaced, or rewritten.
Planning uses temperature/top-$p$ 0.8/0.95, while audio-code decoding uses
0.9/0.95. In the generated-layout condition, the final endpoint of the
validated layout determines both the audio-code target count and renderer
duration. A layout ending at $D$ seconds therefore requests $5D$ audio-code
tokens. Reference- and shuffled-layout conditions likewise use the endpoint of
the supplied effective layout. Systems without a layout representation target
the reference manifest duration.

\subsection{Audio Standardization and Pairing}

Every metric consumes a non-destructively normalized copy of each waveform.
Audio is converted to mono, 44.1-kHz PCM16 and normalized with two-pass
EBU R128~\cite{ebu2023r128}
to $-14$ LUFS with a $-1$ dBTP true-peak ceiling. FAD and CLAPScore use the
complete normalized generated clip. PaSST-KL, SSIM, and acoustic-boundary
agreement operate on filename-matched generated/reference pairs cropped from
time zero to their exact common-minimum duration. No waveform is time-stretched,
looped, or zero-filled to imitate missing musical content.

\subsection{Metric Implementations}

FAD~\cite{roblek2019FAD} compares
VGGish~\cite{hershey2017cnn} embedding distributions.
PaSST-KL~\cite{koutini2022efficient} uses non-overlapping 10-second windows and
averages $\mathrm{KL}(p_{\mathrm{ref}}\Vert p_{\mathrm{gen}})$ across aligned
windows. SSIM~\cite{wang2004image} is computed over paired mel spectrograms.
CLAPScore~\cite{clap} averages prompt--audio cosine similarity over
non-overlapping 10-second windows, zero-padding only the last partial window.

For structural complexity, each clip is represented by the 26-dimensional SCM
descriptor vector~\cite{de2022measuring}. Within a dataset, the reference and
generated vectors are standardized before computing the empirical Energy
Distance~\cite{szekely2013energy}:
\begin{align}
\mathcal{E}(R,G)={}&\frac{2}{nm}\sum_{i=1}^{n}\sum_{j=1}^{m}
\lVert \mathbf r_i-\mathbf g_j\rVert_2 \nonumber\\
&-\frac{1}{n^2}\sum_{i,i'=1}^{n}
\lVert \mathbf r_i-\mathbf r_{i'}\rVert_2 \nonumber\\
&-\frac{1}{m^2}\sum_{j,j'=1}^{m}
\lVert \mathbf g_j-\mathbf g_{j'}\rVert_2.
\end{align}
SCM Energy Distance measures agreement between corpus-level structural
complexity distributions. It is not an item-level measure of whether a system
followed the correct layout.

Acoustic-boundary agreement first extracts time--frequency features, clusters
frames with fixed $k=6$, and obtains a boundary sequence for both reference and
generated audio. The reported $F_{0.5}$ and $F_{3.0}$ use 0.5- and 3-second
tolerances with the standard segment-boundary evaluation
protocol~\cite{turnbull2007supervised}. Unlike SCM, these scores compare
transitions within paired items, although they do not identify which
MusicLayout field caused a boundary.

\section{Descriptive Plan and Output Audits}

The main results measure the realized audio rather than treating a textual
layout as correct merely because it parses. Before audio decoding, we parsed
and schema-validated every sampled layout. An invalid sample was discarded and
sampling was repeated within a finite retry budget, stopping as soon as a valid
layout was obtained. This procedure produced a valid MusicLayout for every
requested evaluation item: all 2,719 FreeMIDI, 1,040 MidiCaps, and 1,000 MuChin
items were successfully validated and rendered. Thus, no evaluation item was
removed because of layout invalidity or rendering failure.
Table~\ref{tab:supp-layout-statistics} summarizes the resulting layout lengths.
these descriptive checks are not substitutes for audio evaluation.

\begin{table}[t]
\centering
\footnotesize
\setlength{\tabcolsep}{1.8pt}
\begin{tabular}{@{}lrrrr@{}}
\toprule
Statistic & \shortstack{FreeMIDI\\train} & \shortstack{FreeMIDI\\generated} &
\shortstack{MidiCaps\\generated} & \shortstack{MuChin\\generated} \\
\midrule
Items & 24,474 & 2,719 & 1,040 & 1,000 \\
Segments & 11.31 & 9.73 & 8.38 & 8.66 \\
Families & 2.28 & 1.67 & 1.50 & 1.50 \\
Recurring families & 1.63 & 1.37 & 1.28 & 1.21 \\
Endpoint (s) & 225.34 & 187.44 & 163.36 & 170.01 \\
\bottomrule
\end{tabular}
\caption{The FreeMIDI train column reports statistics of the MusicLayouts
obtained by applying Algorithm~\ref{alg:musiclayout-extraction} to the FreeMIDI
training split. The other three columns summarize valid MusicLayouts generated
for evaluation. Except for item counts, entries are per-item means. A recurring
family contains at least two segments. Endpoint is the final serialized segment
boundary.}
\label{tab:supp-layout-statistics}
\end{table}

\section{Structural Evaluation and Layout Manipulation}

\subsection{Recurrence Visualization}

The self-similarity matrices in the paper are recurrence visualizations rather
than scalar model-ranking metrics. The analysis uses 12-bin harmonic pitch
class profiles, a 209-ms analysis window, a 139-ms hop, delay-coordinate
embedding over approximately three seconds, and a mutual 4\% nearest-neighbor
recurrence rule following the structure-analysis pipeline of
Serra et al.~\cite{serra2014unsupervised}. Within each target--donor case, all
signals are cropped from time zero to their common minimum duration before SSM
computation. Display intensity is used to make recurrence patterns legible. It
does not enter any quantitative score.

\subsection{Cross-System Recurrence Examples}

\begin{figure}[!t]
\centering
\includegraphics[width=\textwidth]{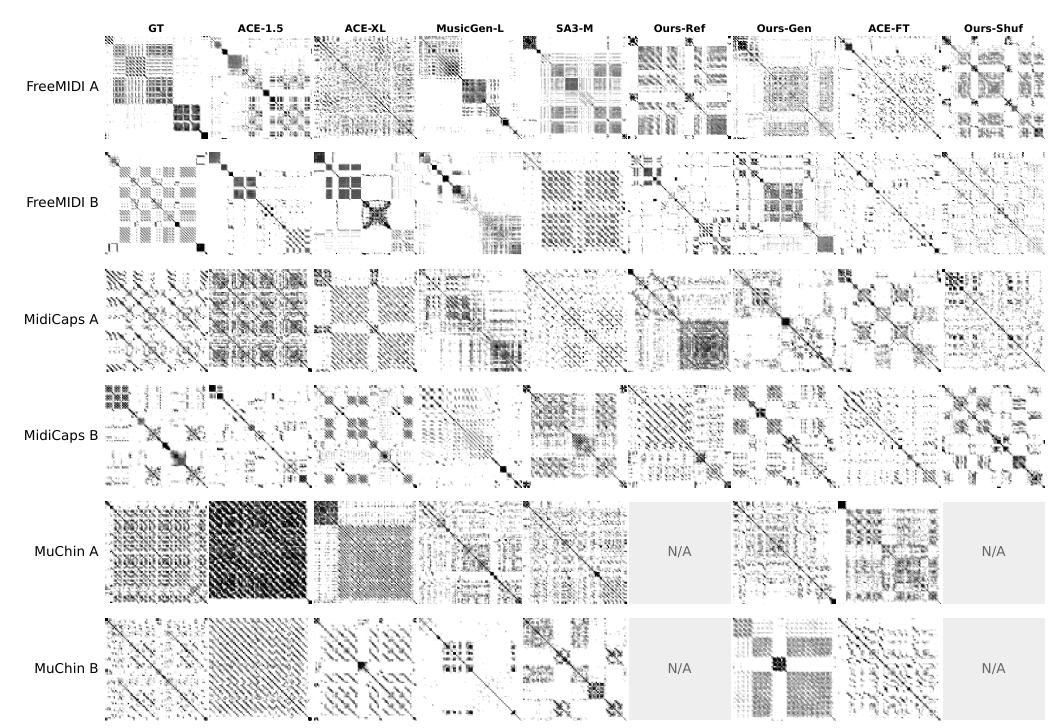}
\caption{Cross-system recurrence SSMs for six cases, with two cases each from
FreeMIDI, MidiCaps, and MuChin. Darker cells indicate
stronger recurrence within a panel. Following the visualization in the main
article, intensities are normalized independently within each panel and therefore
should not be compared as absolute recurrence density across systems. Every row
is cropped to a common duration. Each cell summarizes the mean density of a
$16\times16$ block of the underlying binary recurrence matrix, corresponding
to approximately 2.23 seconds. A monotonic power-law display transform
($\gamma=0.45$) improves the visibility of sparse recurrence patterns without
altering the recurrence matrices. N/A denotes a condition not available for
the MuChin evaluation.}
\label{fig:appendix-all-system-ssm}
\end{figure}

Figure~\ref{fig:appendix-all-system-ssm} expands the recurrence visualization to
the systems in the main objective comparison. It shows two cases from each of
the three evaluation datasets (six rows in total). Each generated example has
the same duration as its corresponding ground-truth audio. Within each row, all available signals
are cropped from time zero to the common minimum duration before recurrence
analysis. The reference-layout and shuffled-layout conditions are unavailable
for MuChin and are marked N/A.

MusicGen-Large primarily exhibits near-diagonal or locally repeated patterns
and less consistently recovers the distant off-diagonal organization visible
in the ground truth. This behavior is consistent with its overlapping-context
generation protocol: a later window receives only the retained tail of the
preceding audio, so patterns from much earlier windows are no longer directly
available as context. Its local continuity therefore does not by itself
preserve long-range recurrence.

On the MIDI-derived rows, the reference-layout condition most closely
preserves the salient block and stripe organization of the corresponding
ground-truth SSMs. This provides qualitative evidence that MusicLayout captures
musically relevant recurrence structure and that the layout-conditioned
renderer can realize that structure in audio. Across the six cases, the
generated-layout condition is less closely aligned with the specific reference,
as expected when the layout is predicted from text rather than extracted from
the target piece, but it still shows repeated blocks and off-diagonal patterns
spanning the generated piece. These examples therefore support reasonable
autonomous long-range organization without treating the SSM visualization as a
scalar model ranking.

\subsection{Cropping Details for Layout Manipulation}

Table~\ref{tab:supp-manipulation-durations} reports the crop durations used for
the four rows of the layout-manipulation SSM in
Figure~\ref{fig:layout_manipulation_ssm}. For each row, the duration is the
minimum available duration among the four signals shown in that row. All
signals are cropped from time zero without temporal alignment, time stretching,
or padding. This operation only establishes a shared visualization interval. It
does not modify the underlying recurrence computation or enter the objective
metrics.

\enlargethispage{2\baselineskip}
\begin{center}
\begin{minipage}{\linewidth}
\centering
\setlength{\abovecaptionskip}{3pt}
\setlength{\belowcaptionskip}{0pt}
\renewcommand{\arraystretch}{0.86}
\footnotesize
\begin{tabular}{@{}lr@{}}
\toprule
Case & Common duration (s) \\
\midrule
FreeMIDI 1 & 224.4 \\
FreeMIDI 2 & 194.6 \\
MidiCaps 1 & 137.4 \\
MidiCaps 2 & 179.716 \\
\bottomrule
\end{tabular}
\captionof{table}{Common-minimum crop durations for the layout-manipulation
SSM rows in Figure~\ref{fig:layout_manipulation_ssm}.}
\label{tab:supp-manipulation-durations}
\end{minipage}
\end{center}

\FloatBarrier

\section{Discussion}

Beyond explicit musical planning, MusicLayout shows how structured
representations can amplify the value of incremental data for music audio
generation. MIDI expands the data available to an audio model, while
extracting layouts further unlocks its temporal and arrangement information.
The resulting prompt--layout--audio triples turn the same incremental corpus
into richer supervision, improving generation while enabling structure-aware
planning and control. Thus, the value of additional data depends not only on
its scale, but also on how its latent structure is represented and exploited.

\section{Qualitative Scope and Limitations}

MusicLayout is a pre-synthesis planning interface. It can change the prefix
that conditions a new waveform, but it does not directly edit an existing
waveform or regenerate a selected region while preserving all other samples.
The categorical instrument representation describes arrangement-level source
classes rather than exact timbres, performances, or production effects. A
valid layout is therefore a well-formed structural request, not a guarantee
that every specified attribute will be realized perfectly in audio.

The model is trained on MIDI-synthesized instrumental audio. This provides
aligned symbolic structure at scale but limits acoustic diversity and fidelity
relative to models trained primarily on studio recordings. It also creates a
domain shift for MuChin, whose evaluation references are accompaniments
separated from real vocal recordings. Consequently, a syntactically valid
layout may still yield audio in which a requested transition, recurrence, or
instrument entry is weak or unclear.

\bibliographystyle{plainnat}
\bibliography{references}

\end{document}